\documentstyle[12pt,lathuile,epsf,times]{article}
\newcommand{\rmt}{\rm\textstyle}
\newcommand{\rms}{\rm\scriptstyle}
\newcommand{\nub}{\overline{\nu}}
\newcommand{\nubar}[0]{\overline{\nu}}

\newcommand{\nunub}{\stackrel{{\footnotesize (-)}}{\nu}}
\newcommand{\txnunub}[1]{\nu_{#1},\nub_{#1}}

\newcommand{\stw}{\mbox{$\sin^2\theta_W$}}
\newcommand{\stwos}{\mbox{$\sin^2\theta_W^{\rms(on-shell)}$}}

\newcommand{\Rnu}{\mbox{$R^{\nu}$}}
\newcommand{\Rnub}{\mbox{$R^{\nub}$}}

\newcommand{\gLeff}{\mbox{$g_L^{\rms eff}$}}
\newcommand{\gReff}{\mbox{$g_R^{\rms eff}$}}
\newcommand{\gLReff}{\mbox{$g_{L,R}^{\rms eff}$}}

\newcommand{\delLReff}{\mbox{$\delta_{L,R}^{\rms eff}$}}
\newcommand{\uLReff}{\mbox{$u_{L,R}^{\rms eff}$}}
\newcommand{\dLReff}{\mbox{$d_{L,R}^{\rms eff}$}}
\newcommand{\qbar}[0]{\overline{q}}

\newcommand{\Ubar}[0]{\overline{U}}

\newcommand{\Sbar}[0]{\overline{S}}
\newcommand{\Dbar}[0]{\overline{D}}
\newcommand{\uubar}[0]{\mbox{$\stackrel{(-)}{u}$}}
\newcommand{\ddbar}[0]{\mbox{$\stackrel{(-)}{d}$}}

\newcommand{\savbar}[0]{\langle\overline{s}(x)\rangle}

\newcommand{\sav}[0]{\langle{s(x)}\rangle}

\begin{document}
\title{ 
A DEPARTURE FROM PREDICTION: ELECTROWEAK PHYSICS AT NUTEV
}
\author{K.~S.~McFarland$^{8,3}$, G.~P.~Zeller$^{5}$,    
 T.~Adams$^{4}$, A.~Alton$^{4}$, S.~Avvakumov$^{8}$, \\
 L.~de~Barbaro$^{5}$, P.~de~Barbaro$^{8}$, R.~H.~Bernstein$^{3}$, 
 A.~Bodek$^{8}$, T.~Bolton$^{4}$,\\ J.~Brau$^{6}$, D.~Buchholz$^{5}$, 
 H.~Budd$^{8}$, L.~Bugel$^{3}$, J.~Conrad$^{2}$,\\ R.~B.~Drucker$^{6}$, 
 B.~T.~Fleming$^{2}$, R.~Frey$^{6}$, J.A.~Formaggio$^{2}$, J.~Goldman$^{4}$,\\ 
 M.~Goncharov$^{4}$, D.~A.~Harris$^{8}$, R.~A.~Johnson$^{1}$, J.~H.~Kim$^{2}$,
 S.~Koutsoliotas$^{2}$, \\M.~J.~Lamm$^{3}$, W.~Marsh$^{3}$, D.~Mason$^{6}$, 
 J.~McDonald$^{7}$, C.~McNulty$^{2}$,\\ 
    D.~Naples$^{7}$, 
 P.~Nienaber$^{3}$, A.~Romosan$^{2}$, W.~K.~Sakumoto$^{8}$, H.~Schellman$^{5}$,\\
 M.~H.~Shaevitz$^{2}$, P.~Spentzouris$^{2}$, E.~G.~Stern$^{2}$, 
 N.~Suwonjandee$^{1}$, M.~Tzanov$^{7}$,\\ M.~Vakili$^{1}$, A.~Vaitaitis$^{2}$, 
 U.~K.~Yang$^{8}$, J.~Yu$^{3}$, and E.~D.~Zimmerman$^{2}$
%
\\
$^1$University of Cincinnati, Cincinnati, OH 45221 \\
$^2$Columbia University, New York, NY 10027 \\
$^3$Fermi National Accelerator Laboratory, Batavia, IL 60510 \\
$^4$Kansas State University, Manhattan, KS 66506 \\
$^5$Northwestern University, Evanston, IL 60208 \\
$^6$University of Oregon, Eugene, OR 97403 \\
$^7$University of Pittsburgh, Pittsburgh, PA 15260 \\
$^8$University of Rochester, Rochester, NY 14627 \\ 
}
%

\maketitle
\baselineskip=14.5pt
\begin{abstract}
  The NuTeV experiment has performed precision measurements of the ratio of
  neutral-current to charged-current cross-sections in high rate, high energy
  neutrino and anti-neutrino beams on a dense, primarily steel, target.  The
  separate neutrino and anti-neutrino beams, high statistics, and improved
  control of other experimental systematics, allow the determination of
  electroweak parameters with significantly greater precision than past $\nu
  N$ scattering experiments.  Our null hypothesis test of the standard model
  prediction measures $\stwos=0.2277\pm0.0013({\rmt stat})\pm0.0009({\rmt
    syst})$, a value which is $3.0\sigma$ above the prediction.  We discuss 
  possible explanations for and implications of this discrepancy.
\end{abstract}
\baselineskip=17pt
\newpage
\section{Introduction and Motivation}

Neutrino scattering played a key role in establishing the structure of the
Standard Model of electroweak unification, and it continues to be one of the
most precise probes of the weak neutral current available experimentally
today.  With the availability of copious data from the production and decay
of on-shell $Z$ and $W$ bosons for comparison, contemporary neutrino
scattering measurements serve to validate the theory over many orders of
magnitude in momentum transfer and provide one of the most precise tests of
the weak couplings of neutrinos.  In addition, precise measurements of weak
interactions far from the boson poles are inherently sensitive to processes
beyond our current knowledge, including possible contributions from
leptoquark and $Z^\prime$ exchange\cite{langacker} and new properties of
neutrinos themselves\cite{oscpaper}.

The Lagrangian for weak neutral current $\nu$--$q$
scattering can be written
as
\begin{eqnarray}
{\cal L}&=&-\frac{G_F\rho_0}{\sqrt{2}}(\nubar\gamma^\mu(1-\gamma^5)\nu)
        \nonumber\\
          &&\times\left( \epsilon^q_L {\qbar}\gamma_\mu(1-\gamma^5){q}+
                \epsilon^q_R {\qbar}\gamma_\mu(1+\gamma^5){q}\right) , 
\label{eqn:lagrangian}
\end{eqnarray}
where deviations from $\rho_0=1$ describe non-standard sources of SU(2)
breaking, and $\epsilon^q_{L,R}$ are the chiral quark couplings~\footnote{Note
  that although we use a process-independent notation here for a tree-level 
  $\rho$, radiative corrections to $\rho$ depend slightly
  on the particles involved in the weak neutral
  interaction.  In this case, $\rho\equiv \sqrt{\rho^{(\nu)}\rho^{(q)}}$.}
For the weak charged current, $\epsilon^q_L=I_{\rms weak}^{(3)}$ and
$\epsilon^q_R=0$, but for the neutral current $\epsilon^q_L$ and
$\epsilon^q_R$ each contain an additional term, $-Q\stw$, where $Q$ is the
quark's electric charge in units of $e$.

The ratio of
neutral current to charged current cross-sections 
for either $\nu$ or $\nub$ scattering
from isoscalar targets of $u$ and $d$ quarks can be written as\cite{llewellyn}
\begin{equation}
R^{\nu(\nub)} \equiv \frac{\sigma(\nunub N\rightarrow\nunub X)}
                 {\sigma(\nunub N\rightarrow\ell^{-(+)}X)}  
= (g_L^2+r^{(-1)}g_R^2),
\label{eqn:ls}
\end{equation}
where
\begin{equation}
r \equiv \frac{\sigma({\overline \nu}N\rightarrow\ell^+X)}
                {\sigma(\nu N\rightarrow\ell^-X)} \sim \frac{1}{2},  
\label{eqn:rdef} 
\end{equation}
and $g_{L,R}^2=(\epsilon^u_{L,R})^2+(\epsilon^d_{L,R})^2$.
Many corrections to Equation~\ref{eqn:ls} are required in a real
target\cite{nc-prl}, but those most uncertain result from the
suppression of the production of charm quarks in the target, which is the
CKM-favored final state for charged-current scattering from the strange sea.
This uncertainty has
limited the precision of previous measurements of electroweak parameters in
neutrino-nucleon scattering\cite{CCFR,CDHS,CHARM}.
One way to reduce the uncertainty on electroweak parameters 
is to measure the observable
 \begin{eqnarray}
R^{-} &\equiv& \frac{\sigma(\nu_{\mu}N\rightarrow\nu_{\mu}X)-
                   \sigma(\nub_{\mu}N\rightarrow\nub_{\mu}X)}
                  {\sigma(\nu_{\mu}N\rightarrow\mu^-X)-  
                   \sigma(\nub_{\mu}N\rightarrow\mu^+X)} \nonumber\\  
&=& \frac{\Rnu-r\Rnub}{1-r}=(g_L^2-g_R^2), 
\label{eqn:rminus}
\end{eqnarray}
first suggested by Paschos and Wolfenstein\cite{Paschos-Wolfenstein}
and valid under the assumption of equal momentum carried by the $u$
and $d$ valence quarks in the target.  Since $\sigma^{\nu
q}=\sigma^{\nub\, \qbar}$ and $\sigma^{\nub q}=\sigma^{\nu \qbar}$,
the effect of scattering from sea quarks, which are symmetric under
the exchange $q\leftrightarrow\qbar$, cancels in the difference of
neutrino and anti-neutrino cross-sections.  Therefore, the suppressed
scattering from the strange sea does not cause large uncertainties in
$R^-$.  $R^-$ is more difficult to measure than $R^\nu$, primarily
because the neutral current scatterings of $\nu$ and $\nub$ yield
identical observed final states which can only be distinguished
through {\em a priori} knowledge of the initial state neutrino.

The experimental details and theoretical treatment of cross-sections
in the NuTeV electroweak measurement are described in detail
elsewhere\cite{nc-prl}. In brief, we measure the experimental ratio of
neutral current to charged current candidates in both a neutrino and
anti-neutrino beam.  A Monte Carlo simulation is used to express these
experimental ratios in terms of fundamental electroweak parameters.
This procedure implicitly corrects for details of the
neutrino cross-sections and experimental backgrounds.  For the
measurement of $\stw$, the sensitivity arises in the $\nu$ beam,
and the measurement in the $\nubar$ beam is the control sample for
systematic uncertainties, as suggested in the Paschos-Wolfenstein
$R^-$ of Eqn.~\ref{eqn:rminus}.  For simulataneous fits to two
electroweak parameters, e.g., $\stw$ and $\rho$ or left and right
handed couplings, this redundant control of systematics cannot be
realized.

\section{Result}\label{sect:results}

As a test of the electroweak predictions for neutrino nucleon scattering,
NuTeV performs a single-parameter fit to $\stw$ with all other parameters
assumed to have their standard values, e.g., standard electroweak radiative
corrections with $\rho_0=1$.   
This fit determines
\begin{eqnarray}
    \sin^2\theta_W^{({\rms on-shell)}}&=&0.22773\pm0.00135({\rmt stat.})\pm0.00093({\rmt syst.})
        \nonumber\\
        &-&0.00022\times(\frac{M_{\rms top}^2-(175 \: \mathrm{GeV})^2}{(50 \: \mathrm{GeV})^2})
        \nonumber\\
        &+&0.00032\times \ln(\frac{M_{\rms Higgs}}{150 \: \mathrm{GeV}}).
\end{eqnarray}
The small dependences in $M_{\rms top}$ and $M_{\rms Higgs}$ result from
radiative corrections as determined from code supplied by
Bardin\cite{bardin} and from V6.34 of ZFITTER\cite{zfitter}; however, it
should be noted that these effects are small given existing constraints on
the top and Higgs masses\cite{LEPEWWG}.  A fit to the precision electroweak
data, excluding neutrino measurements, predicts a value of
$0.2227\pm0.00037$\cite{LEPEWWG,Martin}, approximately $3\sigma$ from the
NuTeV measurement.  In the on-shell scheme, $\sin^2\theta_W\equiv
1-M_{W}^2/M_{Z}^2$, where $M_{W}$ and $M_{Z}$ are the physical gauge boson
masses; therefore, this result implies $M_{W}=80.14\pm0.08$~GeV~\footnote{As
  noted above, this extraction of $\stwos$ is done assuming radiative
  corrections from the standard model as parameterized from $\alpha_{EM}$,
  $G_F$, $M_Z$, $m_{\rms top}$ and $m_{\rms Higgs}$ from fits to the
  electroweak data.  An alternative approach, would be to fit for $M_W$ by
  determining regions of $m_{\rms top}$ and $m_{\rms Higgs}$ favored by the
  NuTeV data and then using those to extract the standard model prediction.}.
The world-average of the direct measurements of $M_W$ is
$80.45\pm0.04$~GeV\cite{LEPEWWG}.  The fact that the NuTeV $\stwos$ deviates
so substantially from $M_W$ makes it difficult to explain the difference
between NuTeV and the standard model prediction in terms of oblique radiative
corrections.

Although NuTeV was primarily designed to measure $\stwos$ using the
Paschos-Wolfenstein relationship, it is also possible to fit for the
single parameter, $\rho_0$.  As noted above, the mechanism by which
the Paschos-Wolfenstein relationship reduces systematic uncertainties
in the $\stw$ fit is evident in the fact that $\Rnu$ only is sensitive
to $\stw$ and thus $\Rnub$ essentially measures systematics common to
the $\nu$ and $\nubar$ beams.  Because $\Rnu$ and $\Rnub$ are both
sensitive to $\rho_0$, there is less control of theoretical
systematics than can be achieved with the $\stw$ measurement, and
uncertainties on $\rho_0$ are therefore larger.  This fit obtains
\begin{eqnarray}
    \rho_0&=&0.9942\pm0.0013({\rmt stat.})\pm0.0016({\rmt syst.})
        \nonumber\\
        &+&0.00006\times(\frac{M_{\rms top}^2-(175 \: \mathrm{GeV})^2}{(50 \: \mathrm{GeV})^2})
        \nonumber\\
        &-&0.00016\times \ln(\frac{M_{\rms Higgs}}{150 \: \mathrm{GeV}}).
\end{eqnarray}
Note that these two results are highly correlated; a simultaneous fit to 
$\stw$ and $\rho_0$ finds:
\begin{equation}
\rho_0=0.99789\pm 0.00405,~\stw=0.22647\pm 0.00311,
\end{equation}
with a correlation coefficient of $0.850$ between the two parameters.  This
suggests one but not both of $\stwos$ or $\rho_0$ may be consistent with
expectations, and that NuTeV is unable to distinguish between these two
possibilities with significant confidence.

Finally, we have also performed a two-parameter fit in terms of the
isoscalar combinations of effective neutral-current quark couplings
$(\gLReff)^2=(\uLReff)^2+(\dLReff)^2$.  The effective couplings, which
describe observed experimental rates when the processes described by
Eqn.~\ref{eqn:lagrangian} are calculated without electroweak radiative
corrections, are measured at $\left< q^2\right> \sim -20$~GeV$^2$.  We
find~\footnote{Note that these coupling results are slightly different
($<<1\sigma$) than the value given in the published NuTeV
result\cite{nc-prl} due to a numerical error.}:
\begin{equation}
(\gLeff)^2=0.30005\pm0.00137,~(\gReff)^2=0.03076\pm0.00110,
\end{equation}
with a correlation coefficient of $-0.017$.  The predicted values from
Standard Model parameters corresponding to the electroweak fit described
earlier\cite{LEPEWWG,Martin} are $(\gLeff)^2=0.3042$ and $(\gReff)^2=0.0301$.
Note that due to the asymmetry between the strange and charm seas and to the
slight excess of neutrons in our target, the NuTeV result is weakly affected
($\sim 1/30$ the sensitivity of $(\gLReff)^2$) by the isovector combinations
of couplings, $(\delLReff)^2=(\uLReff)^2-(\dLReff)^2$.  The results above
assume standard model values for $(\delLReff)^2$.

\section{Interpretation}

The NuTeV $\stw$ result is approximately three standard deviations from the
prediction of the standard electroweak theory.  This, by itself is
surprising; however, it is not immediately apparent what the cause of this
discrepancy might be.  We discuss, in turn, the possibility that the
NuTeV result is a statistical fluctuation among many precision results, the
possibility that unexpected parton asymmetries in the NuTeV affect the
result, and finally possibilities for non-standard physics which could be
appearing in the anomalous NuTeV value.

\subsection{Impact on Standard Model Fits}

\begin{figure}[t]
\mbox{
\epsfxsize=0.47\textwidth\epsfbox{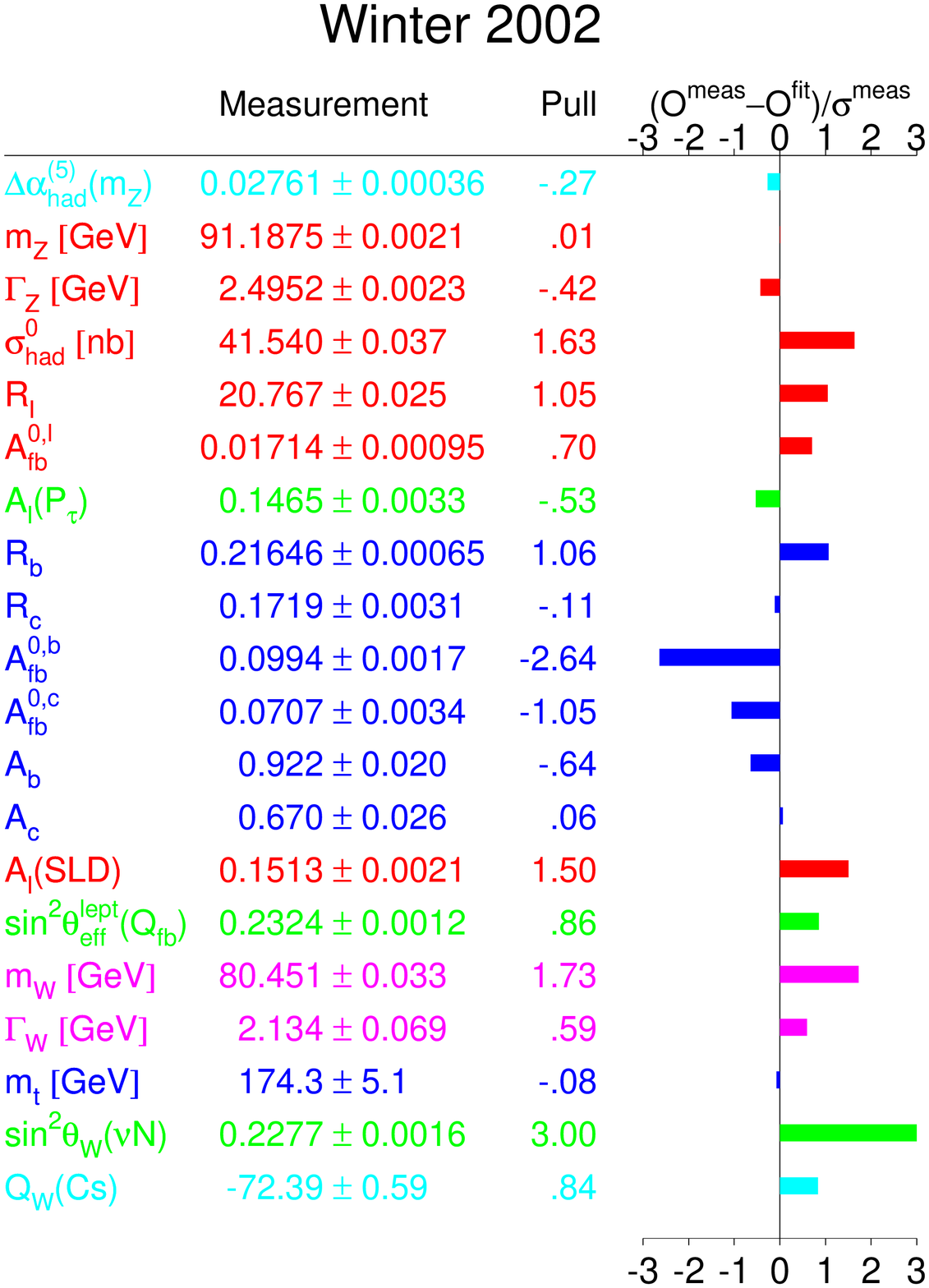}
\raisebox{-0.4cm}{\epsfxsize=0.47\textwidth\epsfbox{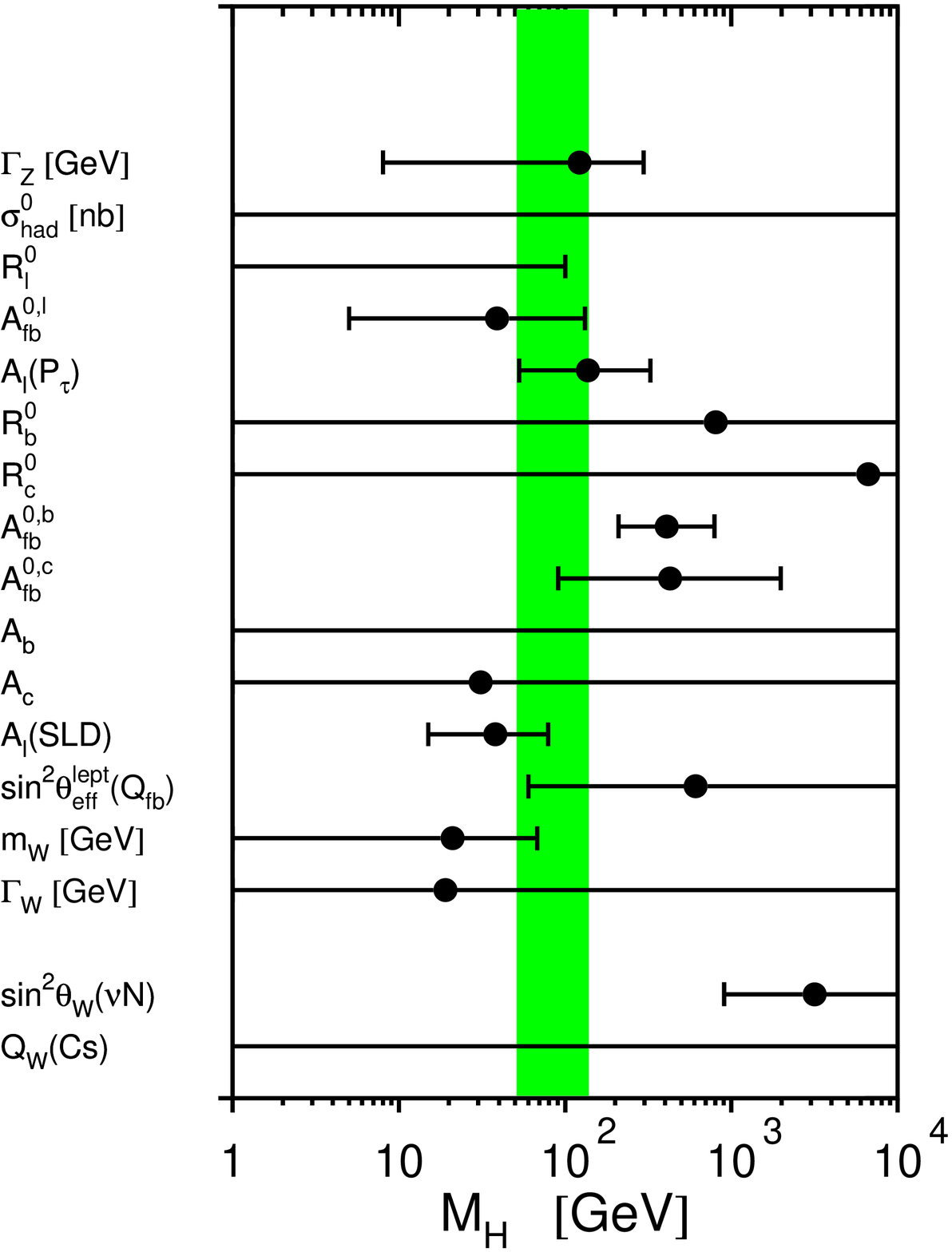}}}
 \caption{\it
      The precision data, as compiled and fit by the LEPEWWG\cite{LEPEWWG}.
      The global fit $\chi^2$ is $28.8/15$.  Most of the data is consistent
      with a low Higgs mass, except $A_{FB}^{0,b}$ and the NuTeV $\stw$.
    \label{fig:LEPEWWG} }
\end{figure}

For fits assuming the validity of the Standard Model, it is appropriate to
consider the {\em a priori} null hypothesis test chosen in the proposal of
the NuTeV experiment, namely the measurement of $\stwos$.  A fit to precision
data~\footnote{It is a tautology, but still worth noting explicitly, that
  certain choices about what constitutes ``the precision data'' must be made
  in order to make such a global analysis.  Some choices that are made in
  compiling this data, for example, not listing the $W$ mass for each
  experiment separately, decrease the number of degrees of freedom without
  significantly decreasing the global $\chi^2$.  Other choices taken, for
  example choosing particular re-evaluations of the central value and
  uncertainties\cite{APV1,APV2,APV3,APV4}, rather than
  others\cite{APV-yetagain} in the atomic parity violation
  measurement\cite{Wieman} of $Q_W^{Cs}$, decrease the global $\chi^2$.},
including NuTeV, has been performed by the LEPEWWG\cite{LEPEWWG}, and the
contribution of each measurement to the $\chi^2$ and final Higgs mass from
this fit is shown in Figure~\ref{fig:LEPEWWG}.  The global $\chi^2$, which
has significant contributions from NuTeV's $\stw$ measurement and
$A_{FB}^{0,b}$ from LEP I, is a rather unhealthy $28.8$ for $15$ degrees of
freedom.  The probability of the fit $\chi^2$ being above $28.8$ is $1.7\%$.
Without NuTeV, this probability of the resulting $\chi^2$ is a plausible
$14\%$.  This suggests that in the context of all the precision data, as
compiled by the LEPEWWG, the NuTeV result is still a statistical anomaly
sufficient to spoil the fit if the standard model is assumed.

This large $\chi^2$ is dominated by two moderately discrepant measurements,
namely $A_{FB}^{0,b}$ and the NuTeV $\stw$, and if one or both are discarded
arbitrarily, then the data is reasonably consistent with the standard model.
However, the procedure of merely discarding one or both of these measurements
to make the fit ``work'' is clearly not rigorous.  Furthermore, the potential
danger of such a procedure has been noted previously in the literature.  For
example, if $A_{FB}^{0,b}$ were disregarded, then the most favored value of
the Higgs mass from the fit would be well below the direct search
limits\cite{Chanowitz}.

Motivated by this large global $\chi^2$ of the precision electroweak
data, we attempt to find other explanations for the discrepancy in the
NuTeV $\stw$ measurement.

\subsection{Unexpected QCD Effects}

As noted above, corrections to Eqns.~\ref{eqn:ls} and \ref{eqn:rminus}
are required to extract electroweak parameters from neutrino
scattering on the NuTeV target.  In particular, these equations assume
targets symmetric under the exchange of $u$ and $d$ quarks, and that
quark seas consist of quarks and anti-quarks with identical momentum
distributions.

The NuTeV analysis corrects for the significant asymmetry of $d$ and $u$
quarks that arises because the NuTeV target, which is primarily composed of
iron, has 
a $5.74\pm0.02$\%
fractional excess of neutrons over protons.
However, this assumption is made under the assumption of isospin symmetry,
i.e., $\uubar_p(x)=\ddbar_n(x)$, $\ddbar_p(x)=\uubar_n(x)$.  This assumption,
if significantly incorrect, could produce a sizable effect in the NuTeV
extraction of $\stw$\cite{Sather,Thomas,Cao,Gambino}.

Dropping the assumptions of symmetric heavy quark seas, isospin symmetry and
a target symmetric in neutrons and protons, but assuming small deviations in
all cases, the effect of these deviations on $R^-$ is\cite{nc-asym}:
\begin{eqnarray}
\delta R^- & \approx 
& - \: \delta N \left( \frac{U_p-\Ubar_p-D_p+\Dbar_p}{U_p-\Ubar_p+D_p-\Dbar_p}\right) (3\Delta_u^2+\Delta_d^2) 
\nonumber\\
&& + \: \frac{(U_p-\Ubar_p-D_n+\Dbar_n)-(D_p-\Dbar_p-U_n+\Ubar_n)}{2(U_p-\Ubar_p+D_p-\Dbar_p)}  (3\Delta_u^2+\Delta_d^2)  \nonumber\\
&& + \: \frac{S_p-\Sbar_p}{U_p-\Ubar_p+D_p-\Dbar_p} (2\Delta_d^2-3(\Delta_d^2+\Delta_u^2)\epsilon_c),  
\label{eqn:deltaR-}
\end{eqnarray}
where $\Delta_{u,d}^2 = (\epsilon^{u,d}_L)^2-(\epsilon^{u,d}_R)^2$, $Q_N$ is
the total momentum carried by quark type $Q$ in nucleon $N$, and the neutron
excess, 
$\delta N \equiv (A-2Z)/A$.  
$\epsilon_c$ denotes the ratio of the
scattering cross section from the strange sea including kinematic suppression
of heavy charm production to that without kinematic suppression.  The first
term is the effect of the neutron excess, which is accounted for in the NuTeV
analysis; the second is the effect of isospin violation and the third is the
effect of an asymmetric strange sea.

\begin{figure}
\begin{center}
\epsfxsize=0.8\textwidth\epsfbox{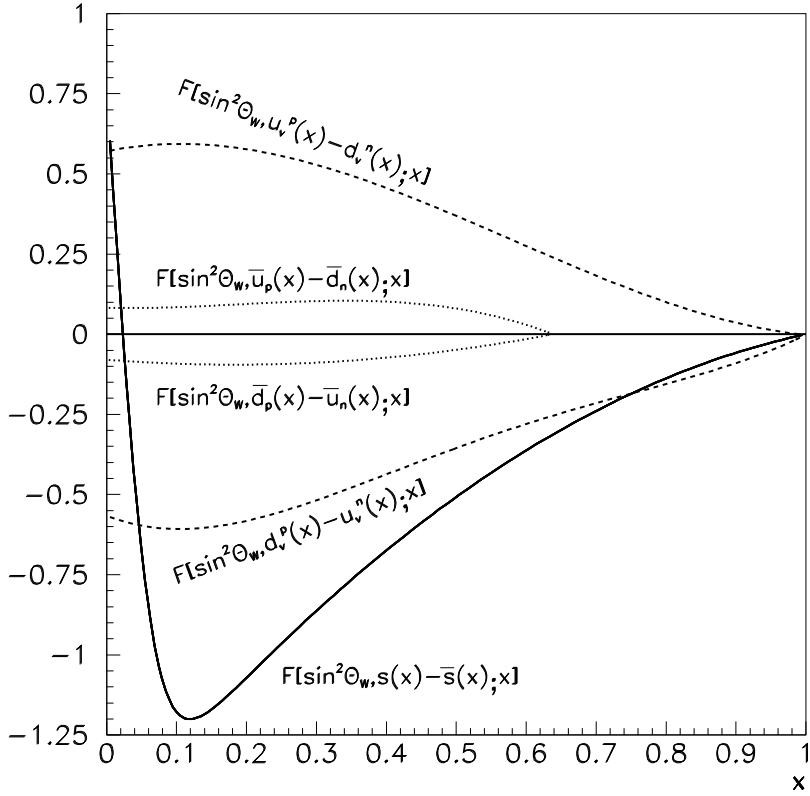}
\end{center}
\caption{\it
  The functionals describing the shift in the NuTeV $\stw$ caused by
  not correcting the NuTeV analysis for isospin violating $u$ and $d$
  valence and sea distributions or for $\sav\neq\savbar$.  The shift in
  $\stw$ is determined by convolving the asymmetric momentum distribution with
  the plotted functional.}
\label{fig:stw}
\end{figure}

NuTeV does not exactly measure $R^-$, in part because
it is not possible experimentally to measure neutral current reactions down
to zero recoil energy.   To parameterize the exact effect of the symmetry
violations above, we define the functional $F[\stw,\delta; x]$ such that 
\begin{equation}
\Delta\stw = \int_0^1 F[\stw,\delta;x]\,\delta(x)\:dx,
\end{equation}
for any symmetry violation $\delta(x)$ in PDFs.  All of the details of the
NuTeV analysis are included in the numerical evaluation of the functionals
shown in Figure~\ref{fig:stw}.  For this analysis, it can be seen that the
level of isospin violation required to shift the $\stw$ measured by NuTeV to
its standard model expectation would be, e.g., $D_p-U_n\sim0.01$ (about 5\%
of $D_p+U_n$), and that the level of asymmetry in the strange sea required
would be $S-\Sbar\sim +0.007$ (about $30\%$ of $S+\Sbar$).

\subsubsection{Isospin Violations}

Several recent classes of non-perturbative models predict isospin violation
in the nucleon\cite{Sather,Thomas,Cao}.  The earliest estimation in the
literature, a bag model calculation\cite{Sather}, predicts large valence
asymmetries of opposite sign in $u_p-d_n$ and $d_p-u_n$ at all $x$, which
would produce a shift in the NuTeV $\stw$ of $-0.0020$.  However, this
estimate neglects a number of effects, and a complete bag model calculation
by Thomas {\em et al.\,}\cite{Thomas} concludes that asymmetries at very
high $x$ are larger, but the asymmetries at moderate $x$ are smaller and of
opposite sign at low $x$, thereby reducing the shift in $\stw$ to a
negligible $-0.0001$.  Finally, the effect is also evaluated in the Meson
Cloud model\cite{Cao}, and there the asymmetries are much smaller at all
$x$, resulting in a modest shift in the NuTeV $\stw$ of $+0.0002$.

Models aside, the NuTeV data itself cannot provide a significant independent
constraint on this form of isospin violation.  However, because PDFs
extracted from neutrino data (on heavy targets) are used to separate sea and
valence quark distributions which affect observables at hadron
colliders\cite{bodek}, global analyses of PDFs including the possibility of
isospin violation may be able to constrain this possibility experimentally.

\subsubsection{Strange Sea Asymmetry}

If the strange sea is generated by purely perturbative QCD processes, then
neglecting electromagnetic effects, one expects $\sav=\savbar$.  However, it
has been noted that non-perturbative QCD effects can generate a significant
momentum asymmetry between the strange and anti-strange
seas\cite{Signal,Burkardt,Brodsky,Melnit}.

By measuring the processes $\txnunub N\to \mu^+\mu^- X$ the CCFR and NuTeV
experiments constrain the difference between the momentum distributions of
the strange and anti-strange seas.
Within the NuTeV cross-section model model, 
this data implies a {\em negative} 
asymmetry\cite{nc-asym},
\begin{equation}
 S-\Sbar = -0.0027 \pm 0.0013,
\end{equation}
\noindent
or an asymmetry of $11\pm6$\% of $(S+\Sbar)$.
Therefore, dropping the assumption of strange-antistrange symmetry results in
an {\em increase} in the NuTeV value of $\stw$,
\begin{equation}
 \Delta\stw = +0.0020 \pm 0.0009.
\end{equation}
\noindent
The initial NuTeV measurement, which assumes $\sav=\savbar$, becomes 
$$\stwos=0.2297\pm0.0019.$$ 
\noindent Hence, if we use the experimental measurement of the
strange sea asymmetry, the discrepancy with the standard model is increased
to $3.7\sigma$ significance.

\subsubsection{Nuclear Shadowing}

A recent comment in the literature\cite{Thomas-Miller} has claimed,
correctly, that if shadowing were significantly different between
charged and neutral current neutrino scattering, this would affect the
NuTeV $\stw$ analysis.  The authors offer a Vector Meson Dominance
(VMD) model of shadowing in which such an effect might arise.  This
model predicts a large enhancement of shadowing at low $Q^2$ which is
not observed in deep inelastic scattering data.  The most precise data
that overlaps the low $x$ and $Q^2$ kinematic region of NuTeV comes
from NMC\cite{NMC-arneodo}, which observed a logarthmic $Q^2$
dependence of the shadowing effect as predicted by perturbative QCD.

Furthermore, shadowing, a low $x$ phenomenon, largely affects the sea quark
distributions which are common between $\nu$ and $\nubar$
cross-sections, and therefore cancel in $R^-$.  The NuTeV
analysis, which uses $\nu$ and $\nubar$ data at $<Q^2>$ of $25$ and
$16$~GeV$^2$, respectively, is far away from the VMD regime, and
therefore the effect even of this VMD model is small, increasing the
prediction for the NuTeV measured $\Rnu$ and $\Rnub$ by $0.6\%$ and
$1.2\%$, respectively.
Finally, the NuTeV $\stw$ data itself disfavors this model through its
separate measurements of $\Rnu$ and $\Rnub$, which are both below
predictions, while this model {\em increases} those very predictions.

\subsection{New Physics}

The primary motivation for embarking on the NuTeV measurement was the
possibility of observing hints of new physics in a precise measurement
of neutrino-nucleon scattering.  NuTeV is well suited as a probe of
non-standard physics for two reasons: first, the precision of the
measurement is a significant improvement, most noticeably in
systematic uncertainties, over previous
measurements\cite{CCFR,CDHS,CHARM}, and second NuTeV's measurement has
unique sensitivity to new processes when compared to other precision
data.  In particular, NuTeV probes weak processes far off-shell, and
thus is sensitive to other tree level processes involving exchanges of
heavy particles.  Also, the initial state particle is a neutrino, and
neutrino couplings are the most poorly constrained by the $Z^0$ pole
data, since they are primarily accessed {\em via} the measurement of
the $Z$ invisible width.

In considering models of new physics, the ``model-independent'' coupling
measurement discussed in Section~\ref{sect:results} is the best guide for
evaluating non-standard contributions to the NuTeV measurements.  An
interesting thing to note about these measurements is that they suggest a
large deviation in the left-handed chiral coupling to the target quarks,
while the right-handed coupling is as expected.  Such a pattern of changes in
couplings is consistent with either a hypothesis of loop corrections that
effect the weak process itself or another tree level contribution that
contributes primarily to the left-handed coupling~\footnote{It has been noted
  many times in the literature that the $A_{FB}^{0,b}$ deviation, combined
  with other constraints on $b$ quark neutral current couplings implies a
  shift in the small {\em right-handed} coupling.  Such a shift is not
  consistent with the hypothesis of a loop-induced correction, either
  standard or non-standard.}.  Chiral coupling deviations are often
parameterized in terms of the mass scale for a unit-coupling ``contact
interaction'' in analogy with the Fermi effectively theory of low-energy weak
interactions.  Assuming a contact interaction described by a Lagrangian of the
form
$$
-{\cal L}=\sum_{H_q\in\{L,R\}} \frac{\pm 4\pi}
{\left( \Lambda^\pm_{LH_q}\right) ^2}\times 
\left\{ \overline{l_L}\gamma^\mu l_{L}\overline{q_{H_q}}\gamma_\mu q_{H_q} 
  + l_{L}\gamma^\mu \overline{l_L}\overline{q_{H_q}}\gamma_\mu q_{H_q}\right.  
\left. + {\rmt C.C.}\right) ,
$$
the NuTeV result can be explained by an interaction with mass scale
$\Lambda_{LL}^+\approx 4\pm 0.8$~TeV.

\subsubsection{Interactions from Extra U(1)}

Phenomenologically, an extra $U(1)$ gauge group which gives rise to
interactions mediated by a heavy $Z^{\prime}$ boson, $m_{Z^{\prime}}\gg m_Z$,
is an attractive model for new physics.  In general, the couplings associated
with this new interaction are arbitrary, although specific models in which a
new $U(1)$ arises may provide predictions or ranges of predictions for these
couplings.  An example of such a model is an $E(6)$ gauge group, which
encompasses the $SU(3)\times SU(2)\times U(1)$ of the standard model and also
predicts several additional $U(1)$ subgroups which lead to observable
interactions mediated by $Z^{\prime}$ bosons\cite{langacker,Cho,Zeppenfeld}.
Before the NuTeV measurement, several authors had suggested in the literature
that the other precision electroweak data favored the possibility of a
$Z^{\prime}$ boson\cite{Casalbuoni,Rosner,Erler,Baur}.

\begin{figure}
\begin{center}
\epsfxsize=0.8\textwidth\epsfbox{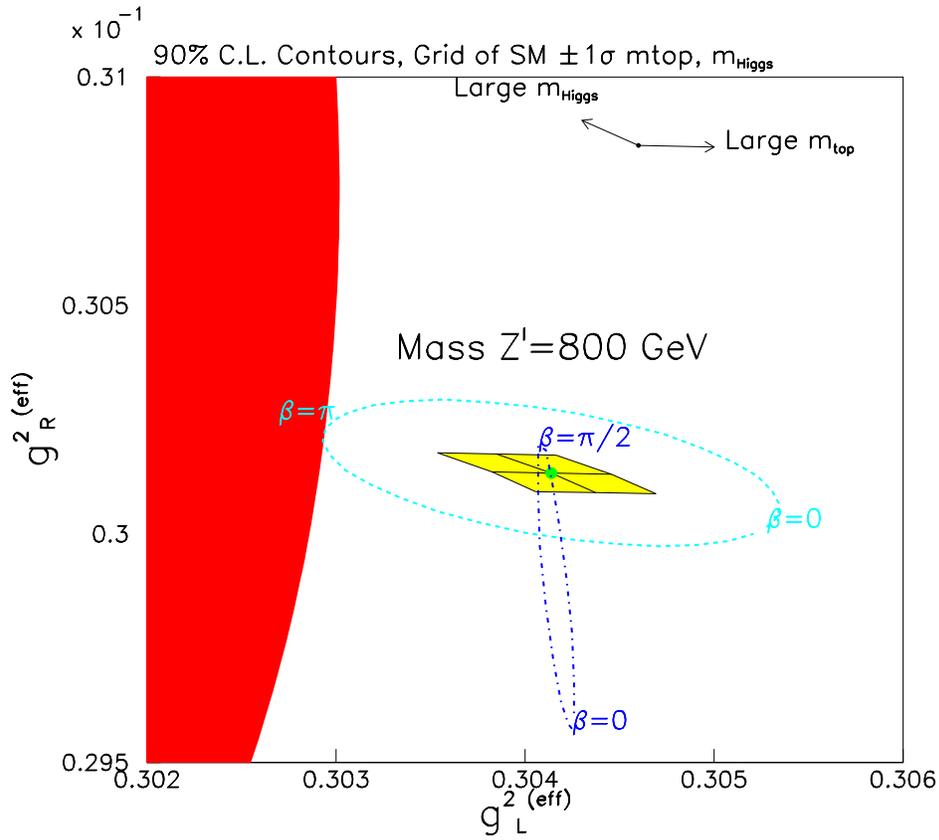}
\end{center}
\caption{\it 
  The effect of $E(6)$ $Z^{\prime}$ bosons on the NuTeV measurement of 
  $(\gLeff)^2$ and $(\gReff)^2$.  The parameter $\beta$ chooses which of the
  possible $U(1)$ subgroups contributed to the observed $Z^{\prime}$.  The
  standard model prediction is the green point, surrounded by a grid of
  $\pm1\sigma$ top and Higgs mass variations.  The upright dark ellipse
  around shows the effect of an unmixed $Z^{\prime}$; the lighter ellipse
  shows the effect of $Z-Z^{\prime}$ a mixing of $0.003$, which is already
  severely constrained by the $Z^0$ pole data.}
\label{fig:e6zprime}
\end{figure}

We have analyzed the effect of $E(6)$-predicted $Z^{\prime}$ bosons on the
NuTeV measurement of the chiral couplings.  As is illustrated in
Figure~\ref{fig:e6zprime}, the effect of these bosons in the case where the
standard model $Z$ and $Z^{\prime}$ do not mix is primarily on the
right-handed coupling.  It is possible to reduce the left-handed coupling
somewhat by allowing $Z-Z^{\prime}$ mixing; however, this possibility is
severely constrained by precision data at the $Z^0$ pole\cite{Erler}.  

More
generally, a $Z^{\prime}$ with couplings of the same magnitude as the $Z$ but
leading to a destructive interference with the $Z$ exchange could explain the
NuTeV measurement if the $Z^{\prime}$ mass were in the range
$\approx1$--$1.5$~TeV.  Current limits from the TeVatron experiments on such
$Z^{\prime}$ are approximately $0.7$~TeV\cite{CDF-zprime, D0-zprime}.
Several authors have also recently discussed other $U(1)$ extensions in the
context of the NuTeV result and found significant effects\cite{Gambino,Ma-Roy}.

\subsubsection{Anomalous Neutrino Neutral Current}

\begin{figure}
\begin{center}
\epsfxsize=\textwidth\epsfbox{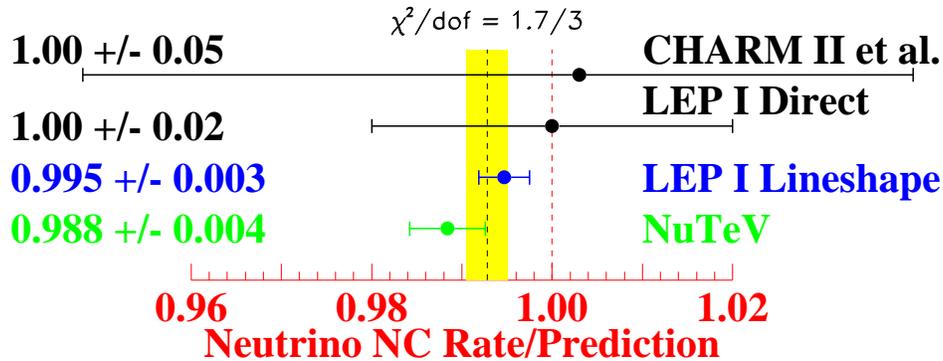}
\end{center}
\caption{\it Measurements of the neutrino current coupling, 
  interpreted as
  a neutrino neutral current interaction rate ($\propto \rho^{(\nu)}$).  The
  precise measurements, $\Gamma (Z\to\nu\nubar)$ at LEP~I and the NuTeV
  measurement of $\rho_0^2$ are both below expectation.}
\label{fig:nunc}
\end{figure}

There are few precision measurements of neutrino neutral current
interactions.  Measurements of neutrino-electron scattering from the CHARM~II
experiment\cite{Charm2} and the direct measurement of $\Gamma(Z\to\nu\nubar)$
from the observation of $Z\to\nu\nubar\gamma$ at the $Z^0$ pole\cite{LEPEWWG}
provide measurements of a few percent precision.  The two most precise
measurements come from the inferred $Z$ invisible width\cite{LEPEWWG} and the
NuTeV result when interpreted as a measurement of $\rho_0^2$ (see
Section~\ref{sect:results}).  As is shown in Figure~\ref{fig:nunc}, both of
the precise rate measurements are significantly below the expectation.
Although this is not a model-independent observation, it is nevertheless
interesting to note this connection between two of the discrepant pieces of
precision electroweak data.

\section{Summary}

The NuTeV experiment has performed a measurement of $\stw$, and finds a
deviation of three standard deviations from the null hypothesis which assumes
the validity of the standard model of electroweak interactions.  Motivated by
the significance of this discrepancy, we study both conventional and new
physics explanations.  Several possibilities exist, although none is
theoretically compelling or has sufficient independent supporting evidence to
be a clear favorite.  Therefore, the cause of this result remains a puzzle.

\section*{Acknowledgements}

We gratefully acknowledge support for this work from the U.S.\ Department of
Energy, the National Science Foundation and the Alfred P.\ Sloan Foundation.
The NuTeV experiment benefitted greatly from significant contributions from
the Fermilab Particle Physics, Computing, Technical and Beams Divisions.  In
addition, we thank Stan Brodsky, Jens Erler, Martin Gr\"{u}newald, 
Paul Langackger, Michael Peskin, Jon Rosner and Tony Thomas for useful 
input and discussions.

\end{document}